\documentstyle[epsfig]{aipproc}
\begin{document}
\title{The Symmetry, Color and Morphology\\ of Galaxies in the Hubble
Deep Field}

\author{Christopher J. Conselice$^*$ and Matthew A. Bershady$^{\dagger}$}
\address{Department of Astronomy, University of Wisconsin-Madison\\
475 N. Charter St. Madison WI, 53706\\
$^*$chris@astro.wisc.edu\\
$^{\dagger}$mab@mingus.astro.wisc.edu}

\maketitle
\vspace{-1cm}
\begin{abstract}

We present a new method of utilizing the color and asymmetry values
for galaxies in the Hubble Deep Field to determine both their morphological
features and physical parameters. By using a color-asymmetry diagram,
we show that various types of star-forming galaxies
(e.g. irregular versus interacting, peculiar galaxies) can be
distinguished in local samples. We apply the same methods to the F814W
images of the Hubble Deep Field, and show preliminary results indicating
that galaxy mergers and interactions are the dominate process responsible
for creating asymmetries in the HDF galaxies.

\end{abstract}

\section*{Introduction}

One of the main purposes of the \it{Hubble Deep Field} \rm{ (HDF) is
to provide an unprecedented opportunity to examine the morphologies of
distant galaxies. However, for a large fraction of the galaxies in the
HDF (40\%), no meaningful morphological indicator of type can be
assigned. 
A problem coupled with the morphology of HDF galaxies is the question
of whether the "peculiar" galaxies are merging systems, or if these
galaxies are just undergoing an intense episode of star-formation. In
this paper, we present a method for determining the morphologies of
galaxies in the HDF based on both galaxy asymmetry and rest-frame UBV
colors. We use a nearby galaxy sample to simulate the appearance of
HDF galaxies, as well as to develop methods that insure reliable comparisons
can be made between nearby galaxies and the more distant HDF
sample. Our preliminary results indicate that a large portion of the
'peculiar' galaxies in the HDF are probably undergoing a merger or
interaction, based on their asymmetry and color values.


\section*{Morphology using Asymmetry}

The first uses of asymmetry for distant galaxy classification
\cite{schade}, \cite{abrah} used asymmetry as a rough morphological
parameter, with image concentration as a second classification
parameter \cite{abrah}. Asymmetry and concentration, while useful for
segregating irregular or 'peculiar' galaxies, do not alone distinguish
between irregular morphology due to interactions, or simply non-uniform
star formation in turbulent environments (e.g.  gas rich, dwarf
irregulars). Conselice \cite{consel97} tested the use of asymmetry as
a general morphological parameter for nearby galaxies, finding a
strong correlation between asymmetry and color.

Here we reformulate earlier methods (\cite{abrah}, \cite{consel97}) to
derive a consistent approach of measuring galaxy asymmetry applicable
at both high and low redshifts. Simulations of nearby galaxies
degraded in resolution and S/N show significant changes in the
asymmetry. It is therefore necessary to correct for these effects,
which are common in high-redshift galaxy images.  In this preliminary
presentation we have corrected for noise, but not for image
degradation. Another critical facet of our new formulation concerns
finding a self-consistent center about which to rotate the galaxy
image. Both \cite{abrah} and \cite{consel97} defined the center of
rotation by the brightest pixel centroid value. This is not a reliable method
for finding the center: galaxy images have brightest pixels which
change considerably with image degradation, yet very small changes in
the center pixel values change the measured asymmetry considerably. To
avoid this problem, we compute the asymmetry for a grid of rotation
points centered on an initial best guess.  We search for the center
yielding the minimum asymmetry value, and iterate the search as
necessary until a true minimum is found.  This allows a robust method
of finding the asymmetry to be computed which is fairly insensitive
(in this regard) to resolution.

For galaxies in the HDF with high S/N $>$ (50) in $I_{814}$ which span
the redshift range from (0$<$z$<$4.5) we compute the asymmetry
parameter using the corrections mentioned above.  We also compute rest
frame B and V colors for each galaxy based on either spectroscopic
redshifts, or from photometric ones based on F300W, F414W, F606W,
F814W, J,H, and K magnitudes \cite{fern}. The $k$-corrections are
computed in the empirical and interpolative manner as described in
\cite{bers}. As such, the rest-frame UBVI colors at higher redshifts,
while accurate, have lower precision because they rely increasingly on
the near-infrared observed bands which are at lower S/N than the
optical data. This should improve with the addition of photometry from
deep NICMOS imaging of the HDF.

\section*{Color-Asymmetry Diagram at $z = 0$}

The color-asymmetry diagram can give a good morphological and physical
indication of the present physical state of a galaxy.  For the nearby
sample of Frei et. al. \cite{frei} there is a strong correlation between the
asymmetry of a galaxy and the color index for non-interacting face-on
systems (Figure 1, left panel). The trend is, as might be expected:
blue galaxies are asymmetric while red galaxies are
symmetric. However, if we plot the entire Frei sample, which contains
galaxies considered 'irregular' or 'peculiar' as well as edge-on disk
galaxy systems, we obtain features which do not lie along the normal
color-asymmetry sequence (Figure 1, labeled sources in left
panel). The fact that these objects do not coincide with the face-on
normal galaxies gives a method for deciphering these objects from
normal face-on high-surface brightness galaxies at high-redshift. In
contrast, the objects at the bluest-asymmetric end of the normal
galaxy sequence are true irregulars -- that is: the asymmetries are
caused by star-formation, and not from projection effects
(inclination), or from interactions.

By simply plotting asymmetry and color for a sample of galaxies, one
cannot immediately disentangle which galaxies falling outside the
normal sequence are inclined or interacting. This requires additional
morphological information obtained by direct image inspection. Since,
for the most part, inclined systems have a high axis ratio, they
can be distinguished easily from interacting systems.

\begin{figure}[b!] 
\epsfig{file=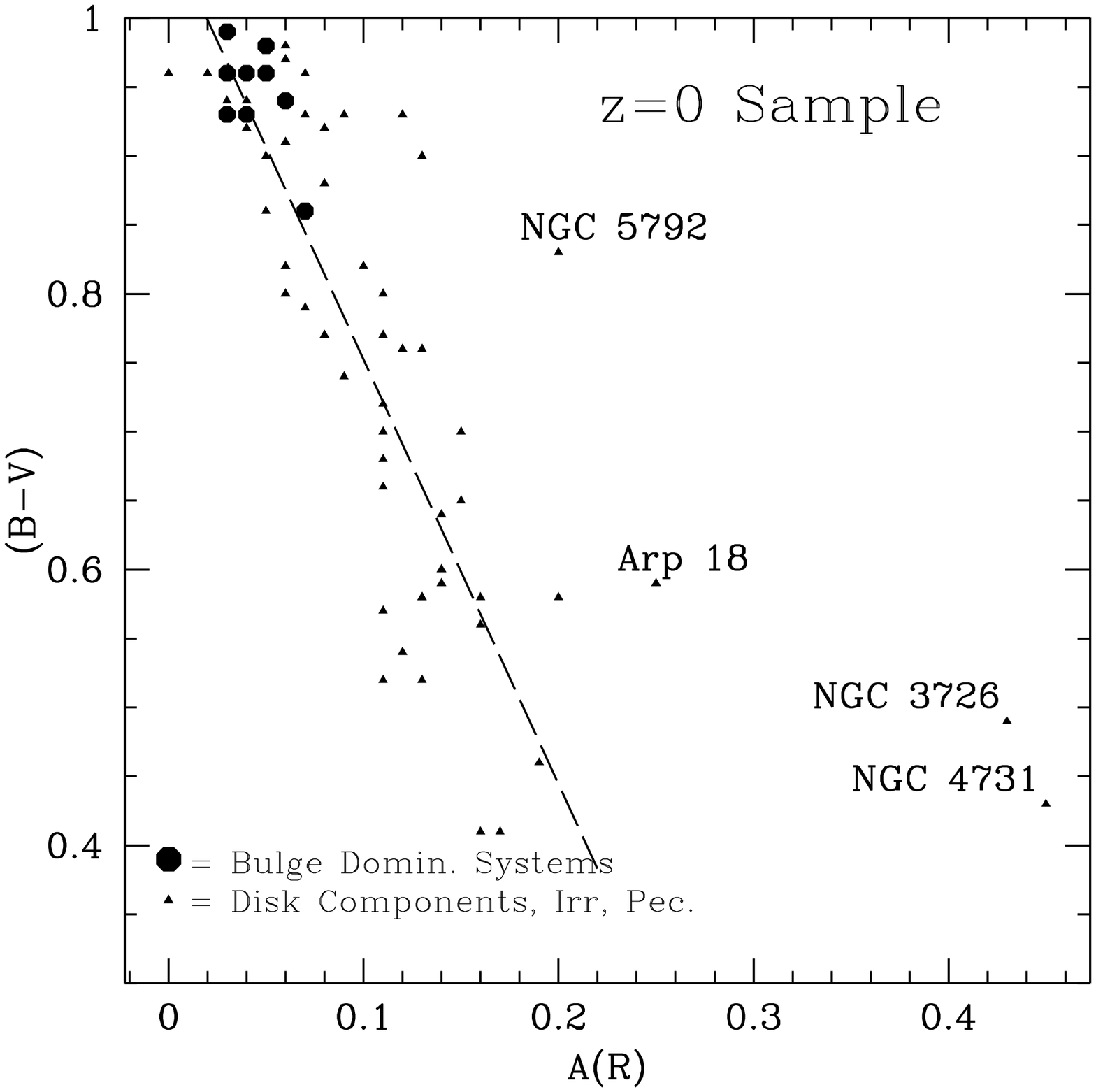,height=2.7in,width=2.7in}
\epsfig{file=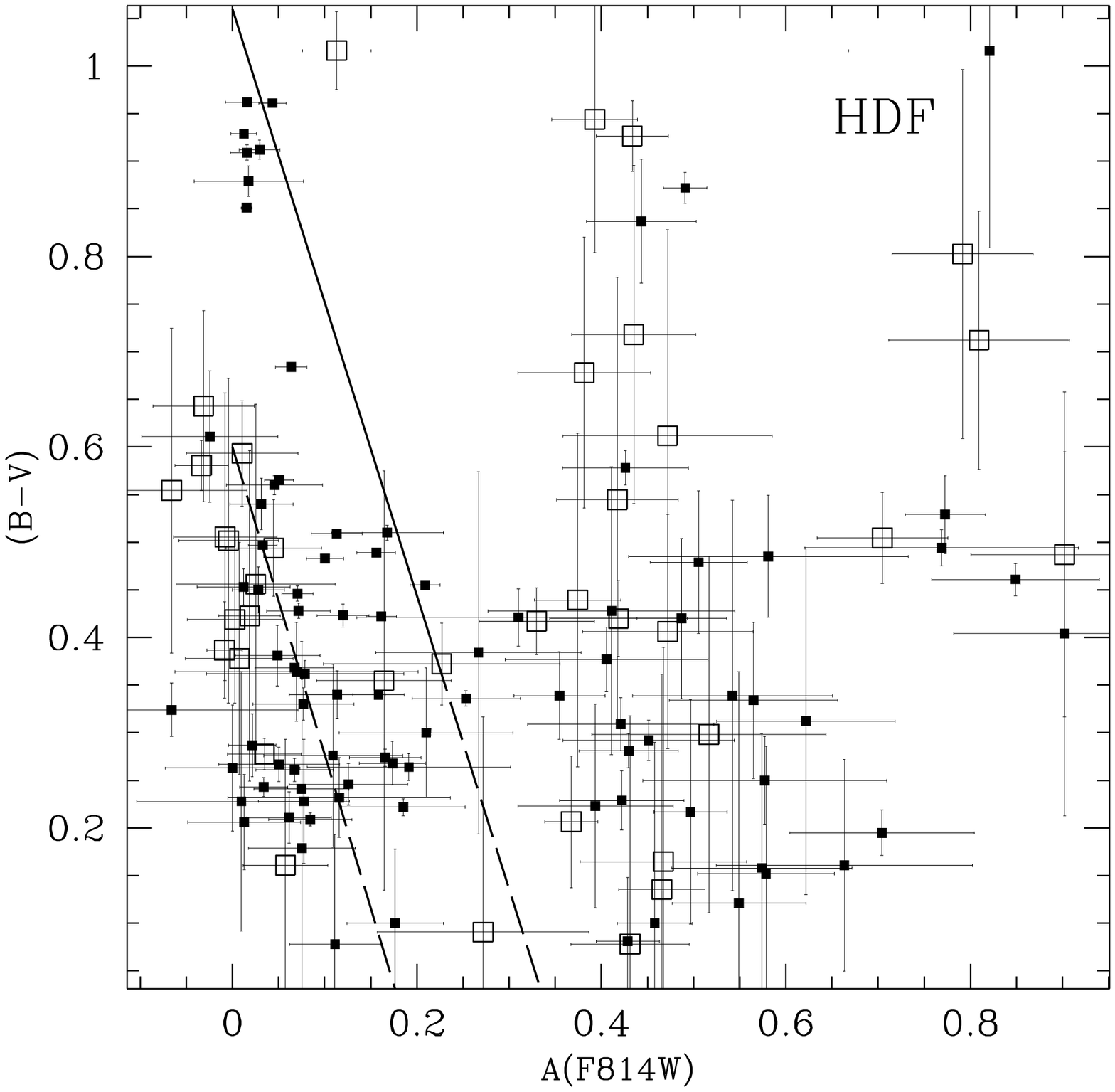,height=2.7in,width=2.7in}
\vspace{10pt}

\caption{Asymmetry-Color diagrams for a local $z=0$ sample (left panel), 
as well as for the high S/N galaxies in the HDF (right pabel). A tight
correlation between color and asymmetry can be seen in the local
sample for most objects with A$<$0.2 (solid line, both panels). The
objects not on this line, labeled with NGC numbers, are galaxies
generally regarded as undergoing an interaction or merger. In the HDF
sample ($z>1$, open symbols; $z<1$, filled symbols), we see a strong
bimodal pattern, with most galaxies either very asymmetric or
symmetric for their blue colors.}
\label{fig1}
\end{figure}

\subsection*{Asymmetries of the Hubble Deep Field galaxies}

To first order the HDF sample {\it avoids} the trend of asymmetry with color
as seen in the local sample of Frei et al.  The majority of galaxies
in the HDF are either too asymmetric or too symmetric for their colors
(compared with the local sample).  However, the local sample includes
almost entirely bright ($\sim$L*) galaxies (presumably as does the HDF
for galaxies at the largest redshifts), while for $z<1$, many of the
HDF galaxies are low luminosity (sub-L*). Indeed, the colors of the
HDF galaxies are blue compared to galaxies in the local Frei et
al. sample, with very few galaxies having colors with (B-V) $>$0.80
(e.g. the colors of un-evolved ellipticals).

It is likely that the highly asymmetric HDF galaxies are the
result of interactions. Normal star-formation processes as defined by
the local sample would still lead galaxies to lie along the local
sequence, albeit at the extreme blue, asymmetric end.  Moreover our
visual inspection of the asymmetric, blue HDF galaxies (A$>$0.3)
reveals that the majority (80\%) are not highly inclined. The
asymmetric 'objects' in the HDF cannot be star-bursting regions
embedded in a largely hidden galaxy, since in this scenario the
bursting region would likely be blue and symmetric.

A class of galaxy not seen in the local sample but present in
profusion in the HDF are blue, symmetric objects. These systems might
be related to the blue-nucleated objects similar to those found in z
$\approx$ 0.5 surveys (e.g.\cite{schade}), although Jangren et
al. \cite{jang} find these specific objects to have A$>$0.2. The
appearance of such a 'new' class of objects could be due either to
increased, nucleated (symmetric) star-formation, or to resolution
effects -- an issue we are exploring.  We emphasize, however, that in
our asymmetry formulation, resolution effects will only tend to lower
A. Hence the highly asymmetric galaxies discussed previously cannot be
an artifact of the analysis.

Our preliminary results indicate that a large portion of the galaxies
in the HDF, while extremely blue, are not undergoing a burst of star
formation in a manner similar to nearby irregular galaxies. For the
most part these galaxies do not appear morphologically as thin, highly
inclined systems, but appear as 'peculiar' galaxies.  Another
substantial fraction of HDF galaxies appear to be highly blue and
symmetric. If this is not a result of decreased physical resolution,
we would surmise these systems have enhanced nuclear
starbursts. Roughly 40\%, however, appear too asymmetric for their
blue color. Such asymmetry is indicative of interactions or mergers
which are disturbing the global light distribution. While these
results are preliminary, if the deviation from the normal-galaxy
color-asymmetry sequence is confirmed to increase with redshift as we
have found, this indicates that {\it merging} is a {\it critical}
process shaping the morphology of high redshift galaxies.

This research was supported by NASA LTSA NAG5-6043 and research funds from 
the UW Graduate School.

\end{document}